# Lost in Math?

J. Butterfield, Trinity College, Cambridge UK: jb56@cam.ac.uk
*Saturday 2 February 2019*

Review for *Physics in Perspective* of:
**Lost in Math: How Beauty Leads Physics Astray,**
by Sabine Hossenfelder. Basic Books 2018.
ISBN: 978-0-465-09425-7, 304 pages, $17.99 (hardcover)

*Abstract:* This is a review of Hossenfelder's book, *Lost in Math: How Beauty Leads Physics Astray*. The book gives a breezy exposition of the present situation in fundamental physics, and raises important questions: both about the content of the physics, and the way physics research is organized. I first state my main disagreements. Then, I mostly praise the book: I concentrate on Hossenfelder's discussion of supersymmetry, naturalness and the multiverse.

### 1. Introduction

This book (Hossenfelder 2018) is an engaging popular account of the present situation in fundamental physics. It is also a strong critique of that situation. All credit to Sabine Hossenfelder for writing a book which---while personal, indeed passionate---is unpretentious and humorous. She is a theoretical physicist in Frankfurt, who also writes an excellent blog about physics at: http://backreaction.blogspot.co.uk. Thus the book is written in a breezy style. It is both an essay on the present situation in fundamental physics, and a memoir of her travels in recent years, conducting extended interviews with about a dozen physicists, including for example: Nima Arkani-Hamed, George Ellis, Gordy Kane, Garrett Lisi, Keith Olive, Joe Polchinski, Steven Weinberg and Frank Wilczek.

Their remarks are reproduced *in extenso*, interweaved with Hossenfelder's thoughts. These are often appealingly ironic and-or self-deprecating. For example, when Polchinski compliments her by saying: 'I think you have tried very hard to separate between ideas that sound like good ideas and ideas that do not seem like good ideas … It's an important thing to do. It's really thankless, though, because the number of bad ideas increases much more rapidly than the number of good ideas …'; Hossenfelder adds: 'This is possibly the nicest way I've ever been told I'm stupid' (p. 177).

In short, the book is engaging, easy to read, and gives vivid explanations of the issues: as I will try to convey with some quotes. And of course, the views of these interviewees are bound to be interesting, especially to readers of this journal.

Hossenfelder's criticisms of the present situation in physics fall into two main groups. The first group concerns the content of recent proposals in fundamental physics; the second, the professional and institutional organization of physics. Hossenfelder devotes most of the book to the first group. Here, she emphasizes supersymmetry, naturalness and the multiverse. She sees all three as wrong turns that physics has made; and as having a common motivation---the pursuit of mathematical beauty. Hence her sub-title. She discusses the second group of criticisms at the end of the book, along with recommendations about how we could, and should, change our ways.

I will follow suit. I will emphasize the first group, especially supersymmetry, naturalness and the multiverse (Sections 3 to 9). I will turn only in the final two Sections (10 and 11) to the organization of physics. But I begin by registering my main disagreements with her (Section 2).

## 2. Disagreements

I am persuaded by much of what Hossenfelder says. In some ways, this is unsurprising. For one main theme of her second group of criticisms is that research in fundamental physics concentrates unduly on a few problems and research programmes, and should instead be more diverse; and it is easy to agree to that. Also, one main theme of her first group of criticisms is that the present difficulties, even defects, of fundamental physics are partly due to a cavalier attitude to some philosophical issues, such as confirmation and explanation. Being a philosopher, I of course find it easy to agree to that.

But as everyone knows: you cannot expect a philosopher (or a book-reviewer) to agree *completely*. Thus I have two main kinds of disagreement: the first is general, the second specifically about beauty.

In general, people are bound to differ in how they weigh the merits---the achievements hitherto, and the future prospects---of a given research programme; even when they are experts with matching knowledge of the programme's technicalities. And so also, as regards comparing the merits of different research programmes. Similarly, for the merits of general 'framework' ideas like naturalness, as against fully-fledged research programmes. And similarly for open problems: people are bound to differ about their merits, i.e. about how fruitful it would be to now address the problem. Thus for all that Hossenfelder says: one undoubtedly could mount a robust and detailed defence of some (perhaps all?) of the programmes and ideas she criticizes.

For example, consider supersymmetry. As we will see, Hossenfelder's main criticism of supersymmetry is, in short, that it is advocated because of its beauty, but is unobserved. But even if supersymmetry is not realized in nature, one might well defend studying it as an invaluable tool for getting a better understanding of quantum field theories.

A similar defence might well be given for studying string theory. In fact, Hossenfelder does not give an extended discussion of string theory. As I said, she focuses on supersymmetry, naturalness and the multiverse; and for the first of these, she stresses supersymmetry in extensions of the standard model, not in string theory. Agreed: she does intermittently mention string theory's lack of empirical confirmation and the beauty of its mathematics; and she briefly quotes practitioners saying, some twenty or thirty years ago, that its beauty suggests it 'had to be pointing to something deep' (thus John Schwarz, on p. 189). But I think many practitioners, especially nowadays, would not motivate string theory by appealing to beauty. Indeed, all agree that it is a very complicated framework with many different types of fields, gauge groups, symmetries etc. (See below on how judgments of beauty are historically conditioned and fallible.) Rather, they would appeal to many technical features: such as how its various intellectual technologies (mostly rooted in quantum field theory) enable one (i) to do many calculations of obvious relevance to quantum gravity, e.g. about black holes, and (ii) to address long-standing issues, such as improving the ultra-violet behaviour of gravity.

Or consider the cosmological multiverse. Here, Hossenfelder's main criticism is, I think, not simply that the multiverse is unobservable: that is, the other pocket universes (domains) apart from our own are unobservable. That is, obviously, 'built in' to the proposal; and so can hardly count as a knock-down objection. The criticism is, rather, that we have very little idea how to confirm a theory postulating such a multiverse. To which the reply will be that, agreed, it is fearfully difficult to do so: but there are various ideas in the literature about how to do so, and so one should just dive in and assess, and try to improve, them. In particular, we should recognize that although some cosmologists are cavalier about what confirmation of a multiverse would amount to: many others show a reflective and nuanced engagement with the question, and related philosophical issues like explanation. So philosophers, and other readers of this book, should resist inferring that the cosmology community as a whole needs, so to speak, to take a course in philosophy of science. There is a wealth of conceptual, indeed philosophical, discussion by practitioners that is not only clear-headed but also inventive.

My specific disagreement is about beauty: i.e. the criticism stated by the sub-title, that the pursuit of mathematical beauty has led physics astray. But for clarity, I should begin with an agreement. Hossenfelder is not concerned with individual, perhaps idiosyncratic, judgments of beauty (or of related ideas, like elegance and simplicity): but rather, with judgments common to a scientific community. And rightly so: it is such communal judgments, not individual ones, that play a significant role in choosing which scientific theory to endorse or to pursue.

This is one reason why it is not helpful to label appeals to beauty as 'subjective'. Recall Kuhn's wise remarks about the perils of this word (1977, pp. 333f.). Incidentally, his essay's main point is that the criteria for theory choice usually invoked---Kuhn lists five: accuracy, broad scope, consistency (both internal and with other theories), simplicity and fruitfulness---cannot be parlayed up into a precise algorithm for choosing theories. For scientists are bound to differ in how to make them precise in any single application, and in how to weigh them when they conflict.  That is surely true. And it is of course the basis of my first, general, disagreement with Hossenfelder: since people are bound to differ, there can be no definitive reasons to reject research programmes, such as string theory, or framework ideas, such as supersymmetry.

But I disagree with Hossenfelder about what the physics community takes the heuristic role of judgments of beauty (and of related ideas, like elegance) to be.

As we will see (Section 4), she argues convincingly that such judgments: (i) are strongly influenced by history, especially by what has been successful in past theories; and (ii) are thoroughly fallible. So I think those who appeal to beauty in physical theorizing must admit these limitations; as must, indeed, anyone who considers the topic.

But here's the rub---how I disagree with Hossenfelder. I also think advocates of beauty as a heuristic *do* admit these limitations. They advocate no more than a historically conditioned, and fallible, heuristic. (Recall my comment that string theorists agree their theory is complicated, and its merits largely technical.) In short, I think Hossenfelder interprets physicists as more gung-ho, more naïve, that beauty is a guide to truth than they really are.

But enough of these cavils. This is a popular book, not an academic one. So it can seem nit-picking for me to emphasize that the programmes and ideas criticized have plenty to say in their defence. Of course they do. And I do not doubt that Hossenfelder, in an academic *modus operandi*, would concede that such defences have weight: though of course, how much weight is a delicate, detailed and inevitably controversial, affair. So having registered my disagreements, I will from now on enter the spirit of the book---starting with how Hossenfelder sketches the present situation in physics.

## 3. Has physics lost its way?

The background to Hossenfelder's accusation, that fundamental physics has lost its way, will be familiar. In the last forty years we have successfully applied the basic frameworks of quantum field theory and general relativity in regimes far beyond those for which they were originally conceived. Here 'far beyond' means several orders of magnitude in some relevant scale such as distance, energy or field-strength. To take just one example from some thirty years ago: recall the 1993 Nobel prize, awarded to Taylor and Hulse. Their analysis of data from a binary pulsar tested general relativity for gravitational fields 10,000 times stronger than the solar system tests---and confirmed it to some ten decimal places. The overall result of this and many similar triumphs is ironic. Namely, that now, these great framework theories, quantum field theory and general relativity, are victims of their own success. For we need to go beyond them, since they face various technical and conceptual problems: such as the hierarchy and cosmological constant problems. But we are in dire need of clues about how to do so.

Thus it is often said (including in this book) that we need data that shows some 'chink in the armour' of these theories: some empirical mismatch that enables us to get a bridgehead for making further progress. And the size of the particle accelerators that now seem to be needed to gather any such data is so enormous as to make them astronomically expensive. Indeed, 'astronomically' can here be understood literally: Hossenfelder mentions (p. 178) that to reach Planckian energies directly (i.e. with the current framework for accelerator design) we would need an accelerator about the size of the Milky Way.

And we are also in dire need of conceptual clues. The standard illustration here is how the null result of the Michelson-Morley experiment presaged special relativity's abandonment of absolute simultaneity. More precisely, to register the fact that the result seems *not* to have influenced Einstein: the null result could have been used as such a clue. So perhaps the present situation is similar. Maybe some empirical facts crucial to solving today's main problems are already to hand, but our current conceptual outlook prevents us recognizing them.

These obstacles, both empirical (financial!) and conceptual, to making progress in fundamental physics, get rehearsed in the book, by both Hossenfelder and her interviewees. I think they would be acknowledged by anyone considering the present state of physics. So a case can be made that in the last forty years, fundamental physics has got stuck.

But even if that is right: it is one thing to be stuck, and another thing to be lost. One can be stuck--- stationary and unable to move---without having taken a wrong turn. Yet Hossenfelder argues that physics has indeed taken a wrong turn. As I announced in Section 1, she makes two kinds of criticism. First: the mentioned obstacles have prompted a mistaken appeal to beauty, and related ideas like simplicity, as a criterion for selecting physical theories. And second: the professional and institutional organization of physics has gone awry. Hossenfelder makes a good, though polemical, case. So it is good news that the book ends with suggestions about how we should mend our ways.

Although the book's interviews and Chapters tend to mingle these two groups of criticisms, the first gets more detailed attention and more pages. And I shall emphasize the first: which is anyway much closer to philosophy, my home ground. So in Sections 4 to 9, I will discuss this first group, emphasizing the three topics: *supersymmetry*, *naturalness*, and the *multiverse*. This group of criticisms leads in to the second, about the social organization of physics: which I discuss in Sections 10 and 11. Finally, I will endorse some of the book's closing suggestions about how we could, and should, 'do better'.

## 4. The dangers of beauty

It is of course sensible, even mandatory, that when we do not have enough evidence to decide between rival theories, we must use other criteria. And there is a strong tradition that physicists should seek beauty and-or simplicity in their equations: a desideratum that some, such as Dirac (p. 25), prized very highly. Besides, in approaching an intractable physics problem, even at a much humbler level than that of Dirac and his ilk, it is often a good idea to seek some elegant or simple formulation. There are countless elementary examples, such as using a symmetry to reduce the number of variables in a problem, and finding normal coordinates.

On the other hand, what counts as beautiful, simple or elegant seems to be in the eye of the beholder, in physics as much as in everyday life. At least, this seems true as regards judgments of these features for general theories as against individual physics problems. That is: such judgments are strongly influenced by history, i.e. by what has been successful in past theories---or past attempts at theories. Indeed, Hossenfelder brings up (in her interviews with Weinberg and Wilczek: pp. 128 and 152) that the philosopher James McAllister (1999) argued that scientific revolutions overturn scientists' conceptions of what counts as beautiful, elegant etc. in their science. Weinberg and Wilczek say they are sympathetic to this view---as they surely should be. A closely related point is that what facts count as needing an explanation is strongly influenced by historical circumstances. An example which Weinberg himself recalls (p. 110) is Kepler's endeavour to explain the relative sizes of the planets' orbits by interpolating Platonic solids between them: sizes that we now accept as historical accidents of how the solar system happens to have evolved, without any explanation in general theory.

Examples like this drive home the lesson that although we might select a theory (or something more preliminary: an attempt at a theory) by its beauty or elegance—especially if we lack evidence for it!---we must in honesty admit that any such judgment of beauty etc. has three great limitations. Namely: (i) it is very much 'by our lights'; (ii) it is fallible; and (iii) it has no logical, or in any way secure, connection with the theory being true. These limitations will recur in the following Sections. In short, the lesson is: beware of the slogan 'too good not to be true'. Beware temptation.

According to the current popular image of fundamental physics, the obvious choice of a target for this warning is string theory, with its lack of empirical confirmation and (alleged!) beauty. Thus it was the target of recent books, which will be familiar to readers of this journal: Smolin's *The Trouble with Physics* (2006) and Woit's *Not Even Wrong* (2006). But as I reported in Section 2, Hossenfelder discusses string theory only briefly. She and her interviewees focus on three other targets---supersymmetry, naturalness and the multiverse: though they each feature in string theory, they also figure prominently outside it.

As to supersymmetry, which is a family of symmetries transposing fermions and bosons: the main point is not merely that it is unobserved. Rather, it is unobserved at the energies recently attained at the LHC at which---one should not say: 'it was predicted to be observed'; but so to speak---'we would have been pleased to see it'. This cautious choice of words reflects the connection to Hossenfelder's second target: naturalness, or in another jargon, *fine-tuning*. More precisely, these labels are each other's opposites: naturalness is, allegedly, a virtue: and fine-tuning is the vice of not being natural.

The main connection between supersymmetry and naturalness, expounded by Hossenfelder, concerns the mass of the Higgs boson. There is intricate advanced physics here, which Hossenfelder explains well in various passages. But before turning to that in Section 7, I will consider naturalness in general terms, in the next two Sections.

## 5. Three kinds of naturalness

Naturalness is, I submit, a cluster of ideas, rather than a single one. I will articulate them as a trio: in order of increasing precision---and perhaps plausibility. Of course, I make no special claim for my trio: one might well explicate and disambiguate 'naturalness' differently. For detailed philosophical discussion, I recommend Williams (2015, especially Sections 3,4; 2018). The second of these papers develops my contrast in (iii) below between typicality as having moderate or higher probability, and as being insensitive to the details of (usually unknown) higher-energy physics. The second paper also relates naturalness to my third theme, the multiverse. In any case: my construal of naturalness as a trio runs as follows.

(i) <u>Against coincidence:</u> *There should be some explanation of the value of a fundamental physical parameter.*
Examples of such parameters are the charge of an electron, or relatedly the fine structure constant. These are traditional examples, owing not least to Eddington's famous---or notorious---speculative efforts (which he called 'Fundamental Theory') to provide such explanations. The contemporary example that Hossenfelder concentrates on (thanks to its connection with SUSY) is the mass of the Higgs boson (cf. Section 5). But whatever the example, the idea here is: the value should not be a 'brute fact', or a 'mere matter of happenstance', or a 'numerical coincidence'.

(ii) <u>Against difference</u>: *The value of a fundamental physical parameter should not be an arithmetical difference of two other physically significant numbers that are nearly equal but both vastly larger in magnitude than the parameter itself.*
      For example: imagine a theoretical framework in which the chosen parameter p is $10^{-6}$. And imagine this is 'because' (i.e. because, according to this framework!) p is the difference of two other numbers, q and r, with some physically significant interpretations, that are nearly equal but both vastly larger than p: e.g. with values $10^6 + 10^{-6}$ and $10^6$. (Of course, as Hossenfelder explains (pp. 64-65): since the value of a parameter usually depends on a human choice of units, the numbers $10^{-6}$ etc. cited here need to be suitably dimensionless.)
      So the framework makes the value of our chosen parameter p *fine-tuned*. It is extremely sensitive to the exact values of these other numbers q and r: in my example, sensitive to their thirteenth digit. Had q and r been slightly different (in terms of proportions of their actual values), then p's value would have been vastly different (proportionately) from its actual value.
      This idea, (ii), is clearly a special case of (i). For the value of p being such a difference is one way for it to be a numerical coincidence. In other words: the presumed theoretical framework, with its equation $p = q - r$, gives only an unsatisfactorily *fragile* derivation of p's value, not a robust explanation of it. And as we will see in Section 7, contemporary physics gives a dramatic example of

this uncomfortable situation. For the mass of the Higgs boson is just such a parameter p: but with the exponent 6 replaced by 14! That is: the mass is a difference of two numbers that, written in decimal notation, match in their first *fourteen* digits, and then differ in the fifteenth digit.

Note finally that this idea, (ii), is also a special case in another regard. For arithmetical difference is of course just one way for our chosen parameter p to be a suspiciously sensitive---a fine-tuned---function of some other numbers. This leads to the more general idea in (iii).

(iii) <u>For typicality</u>: *The value of a fundamental physical parameter should be typical, in some precise sense defined by an appropriate theoretical framework.*

Physics provides two overall schemes for understanding this idea of typicality; Hossenfelder discusses both.

One is general, and has long been endemic in many branches of physics. Namely: there should be a probability distribution over the possible values of the parameter, and the actual value should not have too low a probability. This connects of course with orthodox statistical inference. There, it is standard practice to say that if a probability distribution for some variable is hypothesized, then observing the value of a variable to lie 'in the tail of the distribution'---to have 'a low likelihood' (i.e. low probability, conditional on the hypothesis that the distribution is correct)---disconfirms the hypothesis that the distribution is the correct one: i.e. the hypothesis that the distribution truly governs the variable. This scheme for understanding typicality seems to me, and surely most interested parties---be they physicists or philosophers---sensible, perhaps even mandatory, as part of scientific method. Agreed: questions remain about:

(a) how far under the tail of the distribution---how much of an outlier---an observation can be without disconfirming the hypothesis, i.e. without being 'atypical';

(b) how in general we should understand 'confirm' and 'disconfirm', e.g. whether in Bayesian or in traditional (Neyman-Pearson) terms; and relatedly

(c) whether the probability distribution is subjective or objective; or more generally, what probability really means.

But these questions are not specific to fundamental physics. So I will not pursue them here, and nor does Hossenfelder. But this is not to suggest that they are easy, or that they 'can be safely left to the philosophers'. At the end of this Section, and again in Section 8 (about the multiverse), we shall see questions (a) and (c) return to haunt us.

The second scheme for understanding typicality is specific, and of recent vintage. It developed as part of the effective field theory 'vision' that came from Ken Wilson's deep re-thinking of renormalization group flow. It endeavours to understand a parameter's value being sensitive or insensitive to other parameters' values, in terms of the values' functional dependences---and so apparently without regard to probability. As Hossenfelder explains (pp. 42-48): we envisage a space of theories, where each theory is identified by, roughly speaking, the set of parameters such as coupling constants that occur in its Lagrangian (or Hamiltonian). One then imagines proceeding along a curve in the space of theories, from theories describing high-energy, short-distance, physics to theories describing low-energy, long-distance, physics. This is done, in effect, by integrating out successively more high-energy modes of the fields. Traversing such a curve is called 'following the renormalization group flow'. The question arises: do the curves, along which one proceeds, diverge or converge?

If they diverge, that means that small differences in the parameters of the high-energy theory one started from engender large differences in the parameters of the low-energy theory one arrives at. That is: divergence means the low-energy physics, i.e. the physics we can now observe, is extremely sensitive to the values of parameters describing high-energy physics, i.e. the physics we cannot now, and might never, observe. So divergence means fine-tuning: bad news.  Convergence, on the other hand, would mean that the values of parameters describing low-energy physics are robust to variations in high-energy physics. This suggests the values are 'not a coincidence': good news.

I said that my list, (i) to (iii), would proceed in order of increasing precision, and perhaps plausibility.

It is clear that (i) is vague, and perhaps unpersuasive. The point here is not just that, after all, explanation has to come to an end somewhere. There is also a more specific point. Someone (like myself) who follows David Hume in taking laws of nature to be what Hume called 'constant conjunctions'---i.e. contingent global patterns of co-instantiation of properties---accepts that any explanation, even when it appeals to something as fundamental as laws of nature,  ends in what are ultimately 'just' brute facts, or matters of happenstance. Of course people differ in how easily they accept that explanation must terminate in this sort of way. But to speak for myself, as a Humean: I am content.

Here I can make common cause with Hossenfelder---in part. For this Humean cognitive modesty, this acceptance of limits to a rationalist understanding of nature, meshes well of course with some of what she says in her critique of pursuing beauty. Namely, when she urges (cf. Section 4) that rationalist, even a priori, speculations like Eddington's 'Fundamental Theory' being beautiful (or elegant etc.) gives us no reason to believe them to be true. But this returns us to my second disagreement in Section 2. For I think the great majority of physicists---indeed all right-minded folk!---would concur. They are more modest---less gung-ho, and less naïve---about beauty etc. as a guide to truth, than Hossenfelder portrays them as being. Indeed, one sees this in the book. For sometimes Hossenfelder's interviewees avow scepticism about whether nature must turn out to be rational, or comprehensible, or 'not happenstance': for example, Arkani-Hamed---though he says it with expletives, rather than philosophical jargon (p. 75).

So much for (i). But you should also have misgivings about the more precise proposals, (ii) and (iii). And you should do so, even if you are not a Humean; but instead, full of confidence and ambition about our ability to rationally understand nature.

Thus I noted that (ii)'s main idea, of a suspiciously small vs. a moderate or large difference of real numbers (i.e. a short vs. a long interval of real numbers) leads in to (iii). For (iii) tries to make that idea precise as, essentially: either small vs. moderate-to-large probability, or dependence on small differences at high energy vs. dependence only on large differences at high energy.

But the common theme here---a small vs. large contrast---is all too liable to be a matter of subjective judgment: where (as I said in Section 2) the subjectivity at issue is usually, not of an individual, but of a scientific community. For judgments of small vs. large might well be influenced by our background theoretical beliefs (and thus by our historical circumstances)..

Thus recall questions (a) and (c) in the first, probabilistic, scheme in (iii); and similarly in the second scheme, appealing to the renormalization group flow. Here also, one can---and should---press the point: what counts as a small or large difference in a parameter's value in the high energy regime is very likely to be influenced by theoretical beliefs.

And similarly for naturalness proposals that combine the two schemes (as some do: p. 47) by proposing a probability distribution over parameters' values in the high energy regime, and analyzing its flow down to low energy. One can---and should---press the question: what justifies the measure?

To sum up: I endorse Hossenfelder's summing up of these misgivings: 'the issue of … naturalness shows that we do not understand what it means for a law of nature to be probable' (pp. 208-209). Indeed: we shall see the misgivings above, i.e. the danger that naturalness arguments are infected by subjectivity, in the next Section's historical examples.

## 6. Tycho Brahe and Aristotle

I turn to two examples of arguments based on naturalness, from Renaissance and ancient science. The first is Tycho Brahe's argument against the heliocentric system: Hossenfelder expounds it on pp. 75-77. The second is Aristotle's argument, in *De Anima*, that light cannot be any sort of propagation. Both arguments are plausible, even impressive: though *that* is hardly surprising, given their authors' supreme brilliance.

But both arguments are wrong. For they turn on rejecting as unimaginable, and so unacceptable, a very large value of a physical parameter: in Brahe's case, a distance; and in Aristotle's case, a speed. But so much for human imagination! For in fact, the parameters take those very large values.

Thus Brahe rejected the heliocentric system on the grounds that the stars appear fixed. If the Earth goes round the Sun, then there should be stellar parallax: that is, the stars' apparent positions should change in the course of the year. Indeed they do. But the stars are so distant that the effect was too small for astronomers to measure, until the nineteenth century. After all, our closest star, Proxima Centauri, is 4.2 light-years away: which is about 30,000 times more distant than the planets. However, the idea that the stars are several orders of magnitude further from us than are the planets was, in the sixteenth century, unacceptable. Besides, as Hossenfelder explains, the situation had another confusing feature. Namely: sixteenth century astronomers over-estimated stars' size, so that supposing them to be very far away also implied that they were much larger--- unacceptably larger---than the Sun. Thus she quotes (p. 77) Brahe writing in 1602:

> 'it is necessary to preserve in these matters some decent proportion, lest things reach out to infinity and the just symmetry of creatures and visible things concerning size and distance be abandoned: it is necessary to preserve this symmetry because God, the author of the universe, loves appropriate order, not confusion and disorder'.

Noble sentiments---but wrong.

Similarly two millennia earlier. Aristotle in the *De Anima* argues that light is---not a propagation of anything at all (whether rays or waves or particles), but rather---the transparency of the medium (typically air, of course). He gives as his reason the fact that at sunrise the entire landscape is illuminated at once. Thus he rejects a propagation travelling so fast over the miles-wide landscape, as to look to us as if it arrives everywhere simultaneously. Such a propagation would be `unimaginably fast' in the literal sense: it is unimaginable, and accordingly to be rejected. Thus he writes:

> `Empedocles (and with him all others who used the same forms of expression) was wrong in speaking of light as 'travelling' or being at a given moment between the earth and its envelope, its movement being unobservable by us; that view is contrary both to the clear evidence of argument and to the observed facts; if the distance traversed were short, the movement might have been unobservable, but where the distance is from extreme East to extreme West, the draught upon our powers of belief is too great.' (*De Anima*, Book 2, Chapter 7 (418b21-26: translated by J.A. Smith)

So here we have two arguments whose gist is: 'it must be so, since otherwise some physical parameter would have an unacceptably vast value'. They are both logically valid; and their premises were, at the time, plausible and even compelling. But in fact: the stars are implausibly---'unimaginably'---distant; and the speed of light is unimaginably fast.

The moral is as Hossenfelder stressed above; and I think, as all should agree with; and as we also saw in Weinberg's example (p. 110) of Kepler's trying to explain the relative sizes of the planets' orbits by interpolating Platonic solids. Namely: what we find plausible or acceptable is strongly influenced by our background theoretical beliefs, and thus by our historical circumstances. And so we would do well to interrogate those background beliefs---advice I will return to in Section 9.

## 7. The trouble about the Higgs mass

As I said at the end of Section 4: the main connection between supersymmetry and naturalness, expounded by Hossenfelder, concerns the mass of the Higgs boson.

I shall not linger on the advanced physics of this, which is beyond my ken. (For a philosophical exposition, cf. Williams (2015, Section 2, 3; 2018 Section 2.) I will only quote two passages showing how well Hossenfelder explains the issues. Namely: first, how the Higgs's mass connects supersymmetry and naturalness; and second, what the implied scientific, indeed empirical, problem is. About the first topic, she writes (pp. 37-38):

> 'The Higgs ... suffers from a peculiar mathematical problem that the other elementary particles are immune to: quantum fluctuations make a huge contribution to the Higgs's mass. Contributions like this are normally small, but for the Higgs they lead to a mass much larger than what is observed---too large, indeed, by a factor of $10^{14}$. Not a little bit off, but dramatically, inadmissibly, wrong. ... One can amend the theory by subtracting a term so that the remaining difference gives the mass we observe. This amendment is possible because neither of the terms is separately measurable; only their difference is. Doing this, however, requires that the subtracted term be delicately chosen to almost, but not exactly, cancel the contribution from the quantum fluctuations. For this delicate cancellation one needs a number identical to that generated by the quantum fluctuations for fourteen digits, and then different in the fifteenth digit. But that such a close pair of numbers would come about by chance seems highly unlikely. . . .
> 
> Supersymmetry much improves the situation because it prevents the overly large contributions from quantum fluctuations to the Higgs mass. It does so by enforcing the required delicate cancellations of large contributions without the need to fine-tune. Instead there are only more moderate contributions from the masses of the superpartners. Assuming all masses are natural then implies that the first superpartners should appear not too far away from the Higgs itself. That's because if the superpartners are much heavier than the Higgs, their contribution must be cancelled by a fine-tuned term to give a smaller Higgs mass. And while that is possible, it seems absurd to fine-tune SUSY, since one of the main motivations for SUSY is that it avoids fine-tuning.'

In short: SUSY helps the Higgs' mass to be natural.

So one asks: what is the scientific problem about the Higgs mass? The answer lies in the last part of the quotation just given. Namely: the Higgs has been observed at the LHC, with a mass of 125 GeV---but supersymmetry has not been observed at, or close to, that energy. Nor has it been observed at any energy; (in the LHC's second run, from early 2015 to late 2018, the energy was 13 TeV.)

As Hossenfelder says: '[if] all masses are natural then ... the first superpartners should appear not too far away from the Higgs itself.' Or to quote Gordy Kane's prediction, back in 2001: 'the superpartner masses cannot be very much larger than the Z-boson mass, if this whole approach is valid' (p. 36). This and similar predictions (p. 39) have proven false.

One could of course interpret this situation as illustrating that naturalness, as made precise in the relevant supersymmetric models, has the *merit* of being falsifiable---indeed falsified. But Hossenfelder rejects any such Popperian or falsificationist glee at experiment having killed yet another theory. Instead she concludes (p. 39): 'we know now that if the superpartners exist, their masses must be fine-tuned. Naturalness, it seems, is just not correct. [This] has left particle physicists at a loss for how to proceed. Their most trusted guide appears to have failed them.' Later, she again sums up this scientific, indeed empirical, problem. She writes (pp. 63-64):

> 'The LHC eventually confirmed the Higgs with a mass of 125 GeV, just on the upper edge of the range that had so far not been excluded. A heavier Higgs allows heavier superpartners, so as far as supersymmetry is concerned, the heavier the Higgs, the better. But the fact that no superpartner has yet been found means that such a superpartner would have to be so heavy that the measured Higgs mass could only be achieved by fine-tuning the parameters of the supersymmetric models.'

So Hossenfelder construes the falsification of these specific versions of naturalness as illustrating her main theme. As I summarized it in Section 4 (and endorsed it): judgments of beauty, simplicity etc. are of limited value since they are: (i) historically conditioned (ii) fallible and (iii) without any secure connection to the theory in question being true. Thus she adds, just after this quote, a remark by Keith Olive: 'So now we know there is some fine-tuning. And that itself becomes a very subjective issue. How bad is the fine-tuning?'

## 8. The multiverse

So much by way of discussing supersymmetry and naturalness. I turn to the third main topic of Hossenfelder's critique of the content of current fundamental physics: the multiverse. Hossenfelder's discussion is in two parts: the first part of her interview with Weinberg (pp. 100-116), the end of which covers his influential anthropic prediction of the value of the cosmological constant; and her interview with Ellis (pp. 209-218).

By 'multiverse', I principally mean the multiverse of modern cosmology: not the multiverse of the Everett, or many worlds, interpretation of quantum theory---though I will touch on the latter, here and in Section 9. As we will see, the multiverse is connected with the topics of supersymmetry and naturalness.

Modern cosmology postulates that the very early universe underwent a very brief epoch of accelerating expansion: dubbed 'inflation'. This inflationary epoch preceded the slowing expansion successfully described by the standard Big Bang cosmology that was established in the mid 1970s. (Recall Section 3's discussion of modern physics being a victim of own success.)

This accelerating epoch is very conjectural. It is, however, supported by the fact that it implies features of the cosmic microwave background that are confirmed by observations like that of the COBE and Planck satellites. The epoch is meant to have occurred at times that are (logarithmically!) much earlier---energies much higher, temperatures much hotter---than established physics can describe. Thus it is said to have begun at about $5 \times 10^{-35}$ seconds after the Big Bang, with a characteristic expansion time (i.e. the time-scale on which the scale

factor---the radius of the universe---is multiplied by *e*) of about $10^{-36}$ seconds; and to have ended at about $10^{-34}$ seconds. This is a time corresponding to GUT-scale energies, viz. $10^{15}$ GeV; so that the temperature was $10^{28}$ K. These figures imply expansion by a factor of about $e^{50}$: which is about $10^{22}$!

The putative mechanism driving inflation is yet more conjectural. The main proposed ingredient is a so-far unidentified field, called the inflaton. There are many models, with different details (e.g. potentials for the inflaton). But many such models give rise to a multiverse. That is: to a vast set of (non-interacting) *pocket* or *bubble universes*---of which our own universe is therefore just one. Besides, these bubble universes will vary, not just in details, but also in the values of fundamental physical parameters such as the cosmological constant, and also perhaps the constants governing the strengths of the non-gravitational forces, such as the fine structure constant. This is the *cosmological multiverse*.

Here, there is a confluence with developments in the last twenty years within string theory. Back in the 1980s, string theorists hoped the constraints on constructing a consistent string theory would be so strong that there would be a unique consistent string theory; or at least, only a few such. This hope was ambitious. For the sense of 'theory' at issue here is more specific than a 'framework theory' like, say, elementary classical or quantum mechanics, with their 'silence' or 'generality' about the system's degrees of freedom and the forces impressed on it (i.e. its Hamiltonian or Lagrangian). Here, 'a string theory' is not silent in this sense. It is to encode the forces, i.e. a Hamiltonian; and thereby a ground state.

But this hope has been dashed. In the last twenty years, it has turned out that string theory admits a vast number of local ground states (metastable vacua): states that are in a local minimum of the potential. A truly vast number: an estimate often cited is $10^{500}$, while Taylor and Wang's (2015) estimate is $10^{272,000}$---daunting, indeed depressing, numbers. The 'towers' of excited states built up from each such ground state would then count as different string theories. This is the *string theory landscape.* Besides, the overall scheme of string theory suggests that the different theories---the towers of excited states---differ in values of fundamental physical parameters, such as the cosmological constant and the fine structure constant.

So here is the confluence with the cosmological multiverse. For in both cases, we are confronted with a vast population of what one might (to choose a neutral word) call 'domains': domains that vary in fundamental parameters.

Furthermore, this confluence is strengthened by a widespread, albeit usually implicit, adoption by string theorists of the Everett interpretation of quantum theory. For the idea of the cosmological multiverse is that the countless domains are all *equally real*. I did not stress this above. For one takes it in one's stride, when reading the words 'there is a set of bubble universes, of which our own universe is just one'. But the idea is clear: the domains are 'just' different regions---different 'parishes'---of a vast actuality with an intricate spatiotemporal and causal structure.

But according to the Everett interpretation, one can say *the same* about the various towers of excited states built up from the various vacua in the string theory landscape. That is: suppose that the quantum state of the entire cosmos is some sort of sum, or integral, of states in the various different towers (or a sum of tensor products of such); or is a mixture, i.e. density matrix, with these as components. Then there is, in Everettian parlance, an amplitude for various branches associated with various different vacua: or more precisely, associated with excitations above various different vacua. So there is an amplitude for various different values of fundamental parameters. And according to the Everettian, the different branches are equally real: just as, we saw, the countless domains of the cosmological multiverse are meant to be.

So much by way of describing the dizzying vision of the multiverse, cosmological and string-theoretic. This brief description raises a host of questions, both technical and philosophical. Hossenfelder, with her interviewees Weinberg and Ellis (pp. 100-116, 209-218), explores the latter. As will I, in this Section and the next.
    As I see matters, there are two main philosophical questions to consider, as follows. (For more discussion, cf. e.g. Azhar and Butterfield (2018).) I will take them in turn.
    (1): How, if at all, can we confirm (or disconfirm) a theory postulating such a multiverse, since we can only observe our own domain---our own 'parish'? This question is urgent since it is a treasured hallmark of science that we should be able to confirm (or disconfirm) our claims: a treasure we are in danger of losing. (Cf. Ellis and Silk (2014), who also discuss string theory.) This question will also lead in to the second question.
     (2): What does it take to explain the value of a fundamental physical parameter, such as the cosmological or fine structure constant? Indeed, what does it even *mean* to explain this? In a moment, we will see two more precise versions of this latter question.

As to (1), the obvious approach is to say that a cosmological theory may assign differing probabilities to various values of an observable parameter, of which each domain exhibits one value; and that this enables us to assess the theory by orthodox statistical inference. Namely: the observed value should not be too much of an outlier: not too much 'in the tail' of the probability distribution. That is: if the observed value *is* in the tail, we will conclude that the theory is disconfirmed.
    Admittedly, this approach raises again the usual questions about making precise, and justifying, one's procedures of statistical inference. Recall questions (a) to (c) about the probabilistic construal of naturalness, in (iii) of Section 5.
    But more important: in the present cosmological setting, this approach faces two big problems; (the second in the next Section). Each will lead to a version of question (2) above.

The first problem is: Whence the probability distribution which this approach invokes? It does not help to say 'from our cosmological theory'; or even to say, apparently more specifically, 'from the usual Born-rule probabilities ascribed by a quantum (say string-theoretic) state of the cosmos'. For we do not have such a theory, or such a state, from which we can even in principle calculate a distribution.
    And even if we could calculate one, a nagging question of meaning would remain. For all our understanding of probability derives from cases where there

are many systems (coins, dice … or laboratory systems) that are, or are believed to be, suitably similar. And this puts us at a loss to say what 'the probability of a cosmos', or similar phrases like 'the probability of a state of the universe', or 'the probability of a value of a fundamental parameter' really mean. This is in effect an aggravated version of question (c), about the probabilistic construal of naturalness, in (iii) of Section 5. Namely: 'What is the meaning of probability?' To sum up: we must recognize that Hossenfelder's Section-title, 'Cosmic poker' (pp. 107, 112) is a mere metaphor.

I should add an ancillary remark (which is not in Hossenfelder's discussion). One might say that since all these domains of the multiverse, all these alternative outcomes of the cosmic distribution, are presumed to be actual, we should understand probability in this context *just by counting*. That is: why not take probability to be relative proportion, relative frequency, in the vast actual population of domains? Agreed: in philosophical discussions of probability, identifying probabilities with the corresponding actual relative frequencies (called 'naïve frequentism') is almost always rejected as simplistic. And no doubt rightly: one must allow for stochastic variation---for probabilities not to be exactly matched by the frequencies that happen to occur. But of course, these discussions are not concerned with the cosmos as a whole, nor with a scenario in which all alternative outcomes of the distribution, actually occur---and do so multiply. So perhaps, in this sort of cosmic scenario, 'just counting' is the right way to interpret probability. (This discussion of course bears on the Everett interpretation since, as I mentioned, it is popular among string theorists and, more generally, quantum cosmologists.)

But unfortunately: even if 'just counting' is right, we are really no further ahead. For we still have no idea at all how to calculate any such relative frequencies. The problem is partly that often the sets to be counted are infinite, suggesting a relative frequency of 'infinity/infinity'. And even elementary examples like the proportion of odd numbers among the natural numbers (Guth 2007, p. 6820) show the ambiguities of prescriptions for getting a finite answer.

### 9. Biased sampling

There is also a second problem with what I called 'the obvious approach' of applying statistical inference to the observed value of parameters like the cosmological constant or the fine structure constant.

Namely, the danger of *biased sampling*; or in other jargon, *selection effects*. The issue is familiar in everyday life, and is vividly illustrated by Eddington's (1939) metaphor of the fishing net. Fishermen whose net has a mesh of say 10 centimetres should not infer, from observing that all the fish *in their catch* are longer than 10 centimetres, that the fish *in the lake* are also longer than 10 centimetres.

Similarly for the cosmological multiverse. We observers are like fishermen. Our observations of a physical parameter (e.g. the cosmological constant) are like measurements of the length of fish in the catch. And we should not infer that in unobserved domains---in domains other than ours---the parameter takes, or is likely to take, a value close to what we observe. For maybe the parameter's value is correlated with whether the domain has observers in it.

The point has been well expressed by Hartle, Hertog and Wilczek, as follows. 'What is most probable to occur is not necessarily what is most probable to be observed' (Hartle and Hertog 2017, p. 182); and: 'Observers are located only in places with special properties. As a trivial consequence, probabilities conditioned on the presence of observers will differ grossly from probabilities per unit volume' (Wilczek 2007, p. 43).

In short: Our expectations about what we observe should be conditioned on what we believe about our process of observation.

Agreed: within physics, biased sampling is usually *not* a significant danger. If the observational process is in fact biased---i.e. the value of the parameter we wish to observe is correlated with the process---we can usually recognize this and compensate for the bias. That is: we can quantify the amount of correlation, and accordingly adjust our estimate of the value concerned. But for the observation of fundamental parameters in the context of the multiverse, biassed sampling threatens to be a significant danger---and one that it is very hard to be quantitatively precise about.

The reason lies in the fact that parameters like the cosmological constant or the fine structure constant are correlated with various different aspects of our making observations. Besides, these correlations are often numerically very sensitive. A small change in the parameter would make an enormous change to the observational process, including whether there was an observation at all. Hence the jargon of *fine-tuning;* (meaning, as in Sections 5 and 6, the opposite of naturalness). And furthermore, these correlations often involve various different mechanisms, which are related to one another.

For example, there are many different necessary conditions, each scientifically describable, of our observing the value of the fine structure constant. The observer is alive; and life requires---one may well argue---complex carbon chemistry. Carbon requires stellar nucleosynthesis. And the complex chemistry of life requires---one may argue--- that a planet orbit a star at a suitable distance (neither too hot nor too cold, like Goldilocks' porridge!), and for a long enough time, that life can evolve.

All these correlations, and the mechanisms underpinning them, and these mechanisms' mutual relations, are ferociously hard to disentangle. And this is so even if we somehow settle on some exact definition of 'observation' or 'life'; and it is so even for a single parameter such the fine structure constant, let alone all the physical parameters of interest.

Agreed: these are well-established problems. The *locus classicus* (which Hossenfelder of course cites) is Barrow and Tipler's superb book, *The Anthropic Cosmological Principle* (1986). But they remain as important today as they were thirty years ago, not least because they are made acute and vivid by the cosmological multiverse and the string-theory landscape. Besides, attempts to address them continue.

One influential example, perhaps the most famous example, is Weinberg's explanation of the value of the cosmological constant as an observation selection effect. Despite the complexities just mentioned, this example is comparatively straightforward to calculate: as Weinberg recognized. For the requirement that life evolves in an expanding universe of the ('FRW') type usually considered in cosmology seems to be correlated with the value of the cosmological constant by a single, and comparatively simple, mechanism. Thus Weinberg writes:

> ... in a continually expanding universe, the cosmological constant (unlike charges, masses, etc.) can affect the evolution of life in only one way. Without undue anthropocentrism, it seems safe to assume that in order for any sort of life to arise in an initially homogeneous and isotropic universe, it is necessary for sufficiently large gravitationally bound systems to form first . . . However, once a sufficiently large gravitationally bound system has formed, a cosmological constant would have no further effect on its dynamics, or on the eventual evolution of life (1987, 2607-2608; cf. also 1989, p. 7).

Thus the idea is that the evolution of life constrains the cosmological constant in a simple way, because we can think of (a positive value of) the constant as a long-range repulsive ('anti-gravity') force. Thus one assumes that (i) life can only exist on planets, and (ii) life takes a long time, say billions of years, to evolve. Since (i) requires that matter has the chance to clump together under gravity so as to form planets, the initial expansion cannot be too powerful. That is, there is an upper bound on the cosmological constant. On the other hand, (ii) means that the universe must last long enough for life to evolve. So gravity cannot be so powerful (the initial expansion cannot be so weak) that gravity overcomes the initial expansion in a Big Crunch, well before life has time enough to evolve.

Indeed, the calculation along these lines in 1997, by Weinberg and his co-authors, amounted to a *prediction* that the cosmological constant was positive. (Cf. Martel et al. (1997), which built on previous work, such as Weinberg (1987, 1989 Section V); Vilenkin (2007) is a fine review of the conceptual issues.) The positive value was only measured in the following year: (though there had been earlier hints).

Weinberg, in his interview with Hossenfelder, takes up this example. His comment can serve to summarize our discussion of selection effects. He says (p. 116):

> 'We assumed the probability distribution was completely flat, that all values of the constant are equally likely. Then we said, 'What we see is biased because it has to have a value that allows for the evolution of life. So what is the biased probability distribution?' And we calculated the curve for the probability and asked 'Where is the maximum? What is the most likely value?' ... [Hossenfelder adds: 'the most likely value turned out to be quite close to the value of the cosmological constant which was measured a year later'.] ... So you could say that if you had a fundamental theory that predicted a vast number of individual big bangs with varying values of the dark energy [i.e. cosmological constant] and an intrinsic probability distribution for the cosmological constant that is flat . . . then what living beings would expect to see is exactly what we see.'

The last words convey the lure of what is often called 'anthropic explanation'. But of course, the relevant notion in terms of which we should formulate the process of observation is not humankind, but something more general, such as life, or evolved life. So rather than the label 'anthropic explanation', we should use a phrase like 'parochial explanation', i.e. explanation that invokes the local circumstances of our 'cosmic parish'.

Whatever the label, we face tough questions, about both physics and philosophy. As to physics, I should mention that some recent discussions suggest the string landscape contains *no* universes with a positive cosmological constant. Cf. for example Woit's blog
http://www.math.columbia.edu/~woit/wordpress/?p=10486.
And the philosophical question about explanation is also tough. Can we be content, as Weinberg is, with a parochial explanation of the value of a fundamental physical parameter? Or must we seek an explanation invoking more traditional considerations, especially principles of dynamics and-or symmetry?

## 10. Topics we woefully neglect

So much by way of expounding the multiverse, and its connection with supersymmetry, especially string theory, and naturalness. Together, they form Hossenfelder's three main charges: her three main accusations about how contemporary physics has taken a wrong turn.

Here we should again distinguish, as in Section 3, between a subject getting stuck and it taking a wrong turn. Of course, we can hardly blame the physics community for the subject being stuck in the sense of lacking data. For example: while we would have been pleased to see SUSY in the LHC, it is no one's fault that nature 'chose not to oblige'. Such are the fortunes---oftentimes, misfortunes---of enquiry. No amount of high talent and generous funding guarantees success.

But as we have seen, Hossenfelder argues that the physics community has *erred*, not merely faced misfortune: that physics has taken a wrong turn. She sums up the critique of the previous Sections as follows (pp. 107-108).

> 'The multiverse has gained in popularity while naturalness has come under stress, and physicists now pitch one as the other's alternative. If we can't find a natural explanation for a number, so the argument goes, then there isn't any. Just choosing a parameter is too ugly. Therefore, if the parameter is not natural, then it can take on any value, and for every possible value there's a universe. This leads to the bizarre conclusion that if we don't see supersymmetric particles at the LHC, then we live in a multiverse.
>
> I can't believe what this once-venerable profession has become. Theoretical physicists used to explain what was observed. Now they try to explain why they can't explain what was not observed. And they're not even good at that. In a multiverse, you can't explain the values of parameters; at best you can estimate their likelihood. But there are many ways not to explain something.'

Indeed, a *cri de coeur*. Obviously, the debate continues about whether theoretical physics is in fact in such a parlous state: recall my disagreements in Section 2.

For a ringside seat at the debate, note that the proceedings of the "Why Trust a Theory?" conference at Munich in 2015 are online: https://www.whytrustatheory2015.philosophie.uni-muenchen.de/program/index.html.)
Of course, that conference also addresses several issues I have not mentioned, including philosophical ones: in particular, the important suggestion by Dawid (in his *String Theory and the Scientific Method* (2013: especially Chapter 3) that a theory such as string theory can gather confirmation very indirectly, viz. on account of our failure, despite much effort, to find an alternative theoretical framework.

But even if one thinks Hossenfelder's critique so far is too pessimistic, or unfair, it leads to another: about *neglected problems and research programmes*.

For in the course of her discussions and interviews, Hossenfelder also describes problems that are perfectly valid, even pressing, but are neglected. And correspondingly, for research programmes, rather than problems. There are programmes that are perfectly coherent, even promising, but are unfashionable 'underdogs'. Thus the problems go unaddressed, and the research programmes unfunded.

This critique returns to Section 3's remark that, as well as lacking data at high energies, we are in dire need of conceptual clues about how to go beyond our present theories. Thus recall the analogy with the null result of the Michelson-Morley experiment: maybe some empirical facts crucial to progress today are already to hand, but our mindset prevents us recognizing them.

Of course, people will differ about which such problems and research programmes should be pursued, and which are not worth the candle. But Hossenfelder makes a good case that fundamental physics suffers from a good deal too much conformism, and an unmerited hegemony of a few research programmes. And correlatively, there is far too little collegial and liberal encouragement of diverse approaches.

This critique, this accusation of undue neglect, has two aspects: about the contents of the physics at issue, and about the sociology and professional organization of the discipline. I will take them in turn, in this Section and the next.

Several of Hossenfelder's interviews reveal a neglected problem or an underdog research programme; and some are mostly about such. The obvious example is the interpretation of quantum theory, especially the measurement problem. I say 'obvious', not just because I am a philosopher of physics; nor just because Hossenfelder devotes her Chapter 6 to it. Also: in that Chapter, Weinberg, one of her most distinguished interviewees, is frank and forthright in admitting the problem. For example, he says: 'we don't have any really satisfactory theory of quantum mechanics' (p. 123); and then, in a manner reminiscent of John Bell's famous strictures (e.g. 1989) against invoking the concept of measurement in the postulates of a basic physical theory, he adds (p. 124):

> '... it's a formalism for calculating probabilities that human beings get when they make certain interventions in nature we call experiments. And a theory should not refer to human beings in its postulates. You would like to understand macroscopic things like experimental apparatuses and human beings in terms of the underlying theory. You don't want to see them brought in on the level of axioms of the theory.'

(Weinberg's frank admission of the problem is developed in his 2017 article in *The New York Review of Books*.)

But the interpretation of quantum theory, especially measurement, is by no means Hossenfelder's only neglected issue. For example, she discusses the research programmes of Alain Connes (pp. 157-159) and Xiao-Gang Wen (pp. 190-194). To pick out the latter: the very broad idea is that though physicists are used to the infinities that beset quantum field theory (even the free theory), they are a sickness---and quantum field theory's successes, including the standard model, should be recovered from a thoroughly finitistic basis of qubits.

So much by way of examples. Hossenfelder's broad point about them is that all too often, physicists neglect these problems and research programmes for no good reason. They just follow the all-too fallible guide of aesthetic judgment, or current fashion, or habit; or to put it more neutrally: what the community just happens to get interested in. Thus towards the end of the book, she summarizes with three examples among those I have discussed: the measurement problem, Wen's programme, and the multiverse. She writes (pp. 208-209):

> 'There just isn't any [justification for relying on beauty in selecting theories]. As much as I want to believe that the laws of nature are beautiful, I don't think our sense of beauty is a good guide; in contrast it has distracted us from other, more pressing questions. Like the one that Steven Weinberg pointed out: that we do not understand the emergence of the macroscopic world. Or, as Xiao-Gang Wen reminded me, that we do not understand quantum field theory. Or how the issue of the multiverse and naturalness shows that we do not understand what it means for a law of nature to be probable.'

### 11. How to mend our ways

I turn now to the second aspect of Hossenfelder's accusation of undue neglect. This is about the sociology and professional organization of physics as a discipline.
     Agreed: Hossenfelder's various causes for concern about the organization of physics cannot easily be separated from assessing the content, and confirmation, of physical theories. As every student of science---be they scientist, philosopher, historian or sociologist---will agree: values and attitudes mingle indissolubly with cognition. So one could write a whole review of the book, stressing these concerns. But although they crop up throughout the book, Hossenfelder devotes fewer pages to them; and with space pressing, I will follow suit.

Hossenfelder describes various incidents or patterns of communal behavior, that are undoubtedly peculiar: and most of us would say, dysfunctional---though of course they have their analogues (perhaps worse ones) outside physics, and outside science.

One example is the story of the diphoton anomaly. LHC data from December 2015 suggested a discovery of 'new physics', i.e. incompatibility with the standard model. This was pursued forcefully, also by theorists. For example, ten papers appeared on the arxiv a day after the first announcement; and in the next eight months, five hundred papers were written, some of them getting three hundred citations. But eighteen months later, the anomaly was gone for good (pp. 85, 235). Agreed: in a field starved of new data (Section 3), it is perfectly rational to scrutinize closely a possible anomaly. But given how many problems and research programmes we neglect, it is worrying to see how much effort went into explaining a statistical fluctuation.

Another example is the depressing increase in hype (p. 195). A 2015 survey found that the frequency in scientific papers of words such as 'ground-breaking' and 'unprecedented' increased in the forty years from 1974 to 2014 by a depressing 2500%. (But there is some consolation for physics fans. This survey was of papers in biomedical science . . . so perhaps that explains why nowadays biologists seem to get much more funding than physicists ...)

For Hossenfelder, the main point of such examples is how they reveal considerable cognitive biases in theoretical physicists' selection, or rejection, of problems, and correspondingly, in their support for or scepticism about research programmes. These biases are the main theme of the final Chapter ('Knowledge is power'): which includes a useful summary of some ten main types of bias---useful since the biases described are sadly familiar (pp. 229-231).

> 'There is the sunk cost fallacy, more commonly known as throwing good money after bad ... The in–group bias makes us think researchers in our own field are more intelligent than others . . . the false consensus effect: we tend to overestimate how many other people agree with us and how much they do so . . . Then there is the mother of all biases, the bias blind spot---the insistence that we certainly are not biased.'

So, finally, we face the question: *what can we, what should we, do about it?*

An Appendix, called 'What you can do to help' (p. 245f.), makes some dozen recommendations, aimed at three types of reader of the book. Namely: (i) a scientist; (ii) an administrator in higher education, or science policy maker, or representative of a funding agency; (iii) a science writer or member of the public. All the recommendations are reasonable, and doable: and even if you disagree, they are all worth thinking hard about. Here, to conclude, are some that I for one endorse.

For group (i) i.e. the scientists, the recommendations include: 'Learn about, and try to prevent, social and cognitive biases'. For group (ii), they include: 'Make commitments: you have to get over the idea that all science can be done by post docs on two-year Fellowships . . . short-term funding means short-term thinking'. And: 'Encourage a change of field: if the promise of a research area declines, scientists need a way to get out . . . offer re-education support, one- or two-year grants that allow scientists to learn the basics of a new field'.

And perhaps most uncomfortable of all: For group (iii), the recommendation is: 'Ask questions. You're used to asking about conflicts of interest due to funding from industry. But you should also ask about conflicts of interest due to short-term grants or employment. Does the scientist's future funding depend on producing the results they just told you about? ...'

Here, I submit, Hossenfelder has done physics and science more generally---and our better selves---a service. These are salutary warnings and wise counsels. We should all take notice, and do what we can.

## Acknowledgements
For comments and corrections to previous versions, I am very grateful to: Feraz Azhar, Guido Bacciagaluppi, Alex Chamolly, Richard Dawid, George Ellis, Henrique Gomes, Sabine Hossenfelder, Joe Martin, Sebastien Rivat, Porter Williams, and especially Sebastian De Haro.

## References

Azhar, F. and Butterfield J. (2018). 'Scientific Realism and Primordial Cosmology'; http://arxiv.org/abs/1606.04071; http://philsci-archive.pitt.edu/12192/ abridged in *The Routledge Handbook of Scientific Realism*, ed. J Saatsi; Routledge.

Barrow, J. and Tipler F. (1986). *The Anthropic Cosmological Principle*. Oxford University Press.

Bell, J. (1989). Against measurement. In *62 Years of Uncertainty: Erice 5-14 August 1989*, Plenum Publishers; reprinted in his *Speakable and Unspeakable in Quantum Mechanics*, Cambridge University Press; second edition 2004, pp. 213-231.

Dawid, R. (2013). *String Theory and the Scientific Method*, Cambridge University Press.

Eddington, A. (1939). *The Philosophy of Physical Science*, Cambridge University Press.

Ellis G. and Silk, J. (2014). Defend the integrity of physics. *Nature* **516**, 18-25 December, pp. 321-323.

Guth, A. (2007). Eternal inflation and its implications, *Journal of Physics A: Mathematical and Theoretical* **40**, 6811-6826: hep-th/0702178. doi:10.1088/1751-8113/40/25/S25

Hartle, J. and Hertog, T. (2017). The observer strikes back. In *The Philosophy of Cosmology*, ed. K. Chamcham, J. Silk, J. Barrow and S. Saunders, Cambridge

University Press, pp. 181-205.

Hossenfelder, S. (2018). *Lost in Math: How Beauty Leads Physics Astray,* Basic Books: ISBN: 978-0-465-09425-7, 304 pages, $17.99 (hardcover).

Kuhn, T. (1977). Objectivity, value judgment and theory choice, in his *The Essential Tension*, University of Chicago Press, pp. 320-339.

Martel H., Shapiro P. and Weinberg S. (1997), Likely values of the cosmological constant *Astrophysical Journal* **492**, pp. 29 – 40.

McAllister, J. (1999). *Beauty and Revolution in Science*. Cornell University Press.

Smolin, L. (2006). *The Trouble with Physics*, Houghton Mifflin (Penguin 2007).

Taylor, W and Wang Y-N (2015). The F-theory geometry with most flux vacua, *Journal of High Energy Physics* **12**, pp. 1-21; arxiv.org/abs/1511.03209 doi:10.1007/JHEP12(2015)164

Vilenkin, A. (2007), Anthropic predictions: the case of the cosmological constant, in B. Carr (ed) *Universe or Multiverse?* Cambridge University Press, pp. 163-180. astro-ph/0407586

Weinberg, S. (1987). Anthropic bound on the cosmological constant, *Physical Review Letters*, **59**, 2607 – 2610.

Weinberg, S. (1989). The cosmological constant problem, *Reviews of Modern Physics*, **61**, 1– 23.

S. Weinberg, (2017). The trouble with quantum mechanics, *New York Review of Books*, 19 January 2017; available at:

https://www.nybooks.com/articles/2017/01/19/trouble-with-quantum-mechanics/

Wilczek, F. (2007), Enlightenment, knowledge, ignorance, temptation, in B. Carr (ed) *Universe or Multiverse?* Cambridge University Press, pp. 43-54.

Williams, P. (2015). Naturalness, the autonomy of scales and the 125 GeV Higgs. *Studies in the History and Philosophy of Modern Physics* **51**, pp. 82-96

Williams, P. (2018). Two notions of naturalness. *Foundations of Physics Online* https://doi.org/10.1007/s10701-018-0229-1

Woit, P. (2006) *Not Even Wrong*. Vintage Books.